\documentclass[sigconf]{acmart}
\author{Mohamed Salah Bouafif}
\affiliation{
  \institution{Polytechnique Montréal}
  \city{Montréal}
  \country{Canada}
}
\email{mohamed-salah.bouafif@polymtl.ca}
\author{Mohammad Hamdaqa}
\affiliation{
  \institution{Polytechnique Montréal}
  \city{Montréal}
  \country{Canada}
}
\email{mhamdaqa@polymtl.ca}
\author{Edward Zulkoski}
\affiliation{
  \institution{Quantstamp, Inc}
  \country{United States of America}
}
\email{ed@quantstamp.com}
\AtBeginDocument{%
  }

\setcopyright{acmlicensed}
\copyrightyear{2025}
\acmYear{2025}
\acmDOI{XXXXXXX.XXXXXXX}
\acmConference[EASE 2025]{The 29th International Conference on Evaluation and Assessment in Software Engineering}{17–20 June, 2025}{Istanbul, Türkiye}
\acmISBN{978-1-4503-XXXX-X/2018/06}

\usepackage{algorithm}
\usepackage{algorithmic}
\usepackage{tcolorbox}
\usepackage{tabularx}
\usepackage{lmodern}

\begin{document}

\title{PRIMG : Efficient LLM-driven Test Generation Using Mutant Prioritization }


\renewcommand{\shortauthors}{Bouafif et al.}

\begin{abstract}
Mutation testing is a widely recognized technique for assessing and enhancing the effectiveness of software test suites by introducing deliberate code mutations. However, its application often results in overly large test suites, as developers generate numerous tests to kill specific mutants, increasing computational overhead. This paper introduces PRIMG (Prioritization and Refinement Integrated Mutation-driven Generation), a novel framework for incremental and adaptive test case generation for Solidity smart contracts. PRIMG integrates two core components: a mutation prioritization module, which employs a machine learning model trained on mutant subsumption graphs to predict the usefulness of surviving mutants, and a test case generation module, which utilizes Large Language Models (LLMs) to generate and iteratively refine test cases to achieve syntactic and behavioral correctness.

We evaluated PRIMG on real-world Solidity projects from Code4Arena to assess its effectiveness in improving mutation scores and generating high-quality test cases. The experimental results demonstrate that PRIMG significantly reduces test suite size while maintaining high mutation coverage. The prioritization module consistently outperformed random mutant selection, enabling the generation of high-impact tests with reduced computational effort. Furthermore, the refining process enhanced the correctness and utility of LLM-generated tests, addressing their inherent limitations in handling edge cases and complex program logic.
\end{abstract}


\maketitle

\section{Introduction}
Software testing is a critical component of the software development lifecycle, aimed at ensuring the quality and reliability of software systems. As systems grow in complexity, the need for effective and efficient testing strategies becomes increasingly paramount. One of the primary challenges in developing efficient test suites is achieving comprehensive coverage while minimizing redundancy. As software systems become more complex, the number of possible execution paths and scenarios increases exponentially. This makes it difficult to create test cases that adequately cover all aspects of the system without becoming unwieldy or time-consuming to execute~\cite{mcminn2011search}.
To address these challenges, researchers and practitioners have turned to advanced techniques such as search-based software testing (SBST) and heuristic approaches. SBST applies optimization algorithms, such as genetic algorithms, to automatically generate test cases that maximize coverage and fault detection while minimizing the overall size of the test suite. These techniques can explore vast search spaces more efficiently than manual approaches, potentially uncovering edge cases and vulnerabilities that might otherwise be missed. While such approaches have proven their efficiency in generating test cases that uncover edge cases, they require significant computational resources, especially for large and complex systems.

Mutation testing (MT) has emerged as a key technique for evaluating and enhancing test suite quality by introducing faulty program variants, known as mutants, to assess error detection capabilities. Beyond measuring test effectiveness, recent approaches leverage surviving mutants to guide new test case generation \cite{tip2024llmorpheus, du2024leveraging}. However, MT can be computationally intensive due to the large number of mutants and the potential growth of the test suite. To address this, Liu et al. proposed AID \cite{liu2024llm}, combining large language models (LLMs) with differential testing to generate fault-revealing tests for programs that pass traditional suites. Despite these advances, challenges remain in generating correct and valid test cases. Taherkhani et al.~\cite{taherkhani2024valtestautomatedvalidationlanguage} highlighted accuracy issues in LLM-generated tests, while Yuan et al.'s ChatTester \cite{yuan2023no} iteratively refines test cases to enhance quality, underscoring the need for robust validation mechanisms.

While significant efforts have explored the use of Large Language Models (LLMs) for test case generation, several challenges have emerged, primarily related to ensuring the correctness of the generated tests and the computational intensity required to construct high-coverage test suites. Mutation testing aims to improve test quality and coverage by guiding developers to create tests targeting surviving mutants. These surviving mutants serve as test goals, with the objective of enhancing code coverage and identifying hidden vulnerabilities. However, as the complexity of software systems grows, the number of generated mutants increases exponentially, leading to scalability challenges in the process of test generation. Specifically, generating and validating test cases to kill a large number of mutants becomes a computationally expensive and time-consuming process, even when employing LLMs, which require additional resources for iterative refinement. Despite these scalability concerns, the state-of-the-art mutant selection technique remains random selection, which often fails to optimize computational efforts and test effectiveness \cite{kurtz2016analyzing, gopinath2016limits}. 

In this paper, we introduce PRIMG (Prioritization and Refinement Integrated Mutation-driven Generation), a novel approach tailored for incremental test suite generation. PRIMG leverages mutation testing to guide the effective use of LLMs in generating and refining test cases for Solidity smart contracts. Our approach aims to achieve effective test generation that optimizes the total number of generated tests while simultaneously enhancing the mutation score.  Our proposed technique, employs two main components: a \textbf{mutation prioritization module}, responsible for selecting mutants as test goals for the generated test cases, and a \textbf{test case generation module}, which employs an LLM for generating and refining the test cases.

The rest of this paper is organized as follows. Section 2 briefly reviews the related works. Section 3 describes the different modules of our approach. We present our experimental setup, research questions, and experimental results in Section 4. We discuss our findings in Section 5. Threats to the validity of our results are reviewed in Section 6. Finally, we conclude the paper in Section 7, highlighting some avenues for future work.

\section{Related Works}
\label{sec:motivation}

The proliferation of smart contracts has established them as a cornerstone of automation and trust, facilitating secure and transparent transactions across diverse industries, including finance, real estate, and supply chain management~\cite{de2019old}. However, their increasing adoption has unveiled numerous challenges. The immutable and transparent nature of smart contracts paradoxically renders them susceptible to exploitation, misuse, and unforeseen vulnerabilities \cite{zhou2023sok}. Consequently, the imperative to conduct rigorous audits of smart contracts arises, entailing a meticulous examination of their code and features to detect and mitigate potential risks\cite{zhou2023sok}. The immutable nature of smart contracts adds a new layer of importance for developers and enhances the need for the development of a comprehensive and complete test suite~\cite{gurbuz2024test}.

Over the years, various tools have been developed to automate the task of test case generation for smart contracts. Numerous studies have explored the use of genetic algorithms for test case generation~\cite{ji2022test, driessen2021automated}. For example, the authors of~\cite{olsthoorn2022syntest} investigated the application of meta-heuristic algorithms, implementing multiple search techniques such as random search and genetic algorithms (e.g., NSGAII\cite{olsthoorn2022syntest}, MOSA\cite{olsthoorn2022syntest}, and DynaMOSA\cite{Panichella2018AutomatedTC}) for effective test case generation. While these approaches have demonstrated promising results, the genetic algorithm approach remains computationally expensive.

As highlighted in the in-depth study by Zou et al.~\cite{8847638}, generating test suites for smart contracts is a complex task, with several challenges arising throughout their development and implementation. Among the surveyed developers, 54.7\% reported the lack of mature and business-ready tools for blockchain-specific development. Additionally, these tools fail to address all corner cases and scenarios, which is the most critical challenge identified by the developers.

The emergence of large language models (LLMs) for software development tasks \cite{hou2023large} has sparked interest in leveraging these models for test case generation \cite{ouedraogo2024large}. Several tools and techniques have been proposed in this area. For example, Arghavan Moradi Dakhela et al.~\cite{dakhel2024effective} suggested using mutation testing to guide LLMs in generating effective test cases. After an initial mutation campaign, their approach prompts the LLM with surviving mutants, requesting the generation of tests to kill the mutants. 
While their empirical evaluation showed promising results, the approach has yet to be tested on large datasets and real-world projects. Large codebases introduce three significant challenges: (1) the large number of generated mutants makes performing mutation testing computationally expensive and time-consuming, (2) the large number of tests required to achieve a high mutation score, and (3) real-world project requires intricate test setups and environments and more complex test case then simple assert-statement tests.

\section{Approach}
\label{sec:approach}

\begin{figure*}[t!]
\centering
\includegraphics[width=.9\textwidth]{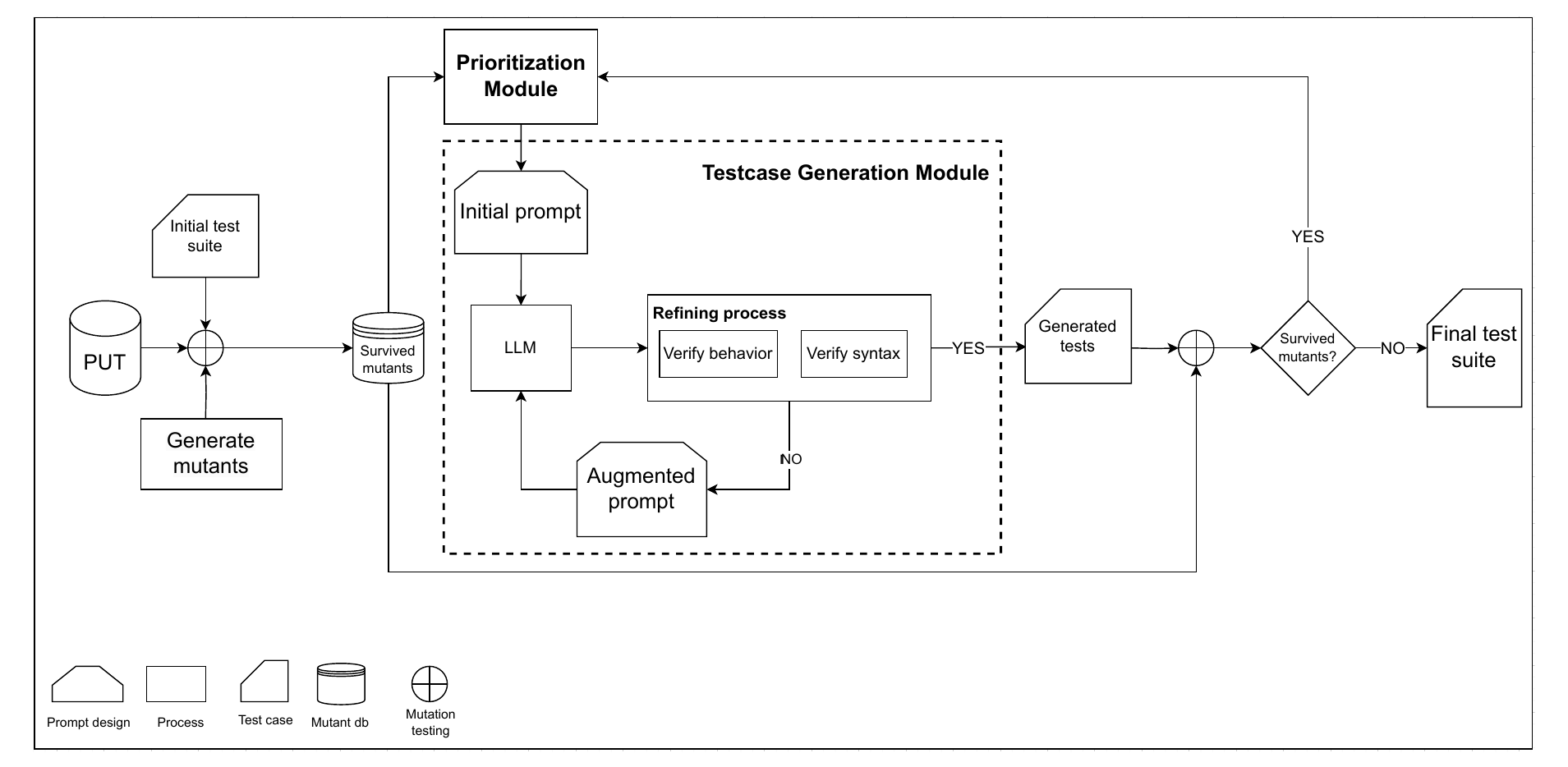}
\caption{The proposed approach for generating a test case using LLMs.}
\label{WFM}
\Description{}

\end{figure*}

In this section, we discuss the different steps of our approach. Figure \ref{WFM} shows an overview of our proposed approach, and Algorithm \ref{alg:PRIMG} presents the sequence of its different steps.
PRIMG is composed of two main modules: \textbf{testcase generation module} and \textbf{mutants prioritization module}. We discuss each of the two modules in the following subsections.

\begin{algorithm}
\caption{PRIMG}
\label{alg:PRIMG}
\textbf{Input:} \textit{PUT}, \textit{initialTestSuite}, \textit{survivingMutants} \\
\textbf{Output:} \textit{finalTestSuite} \\

\begin{algorithmic}[1]
    \STATE \textit{prioritizedMutant} $\leftarrow$ \textbf{PrioritizeMutant}(\textit{survivingMutants})
    \STATE \textit{initTest} $\leftarrow$ \textbf{SelectTest}(\textit{prioritizedMutant}, \textit{initialTestSuite})
    \STATE \textit{initPrompt} $\leftarrow$ \textbf{CONCAT}(\textit{initTest}, \textit{prioritizedMutant}, \textit{PUT})
    \STATE \textit{rawIUT} $\leftarrow$ \textbf{LLM}(\textit{initPrompt})
    \STATE \textit{FUT}, \textit{success} $\leftarrow$ \textbf{RefiningTest}(\textit{rawIUT}, \textit{PUT}, \textit{prioritizedMutant})
    \IF{ \textit{sucess}}
        \STATE Add \textit{FUT} to \textit{finalTestSuite}
    \ENDIF

\STATE \textbf{return} \textit{finalTestSuite}
\end{algorithmic}
\end{algorithm}

\begin{algorithm}
\caption{RefiningTest}
\label{alg:RefiningTest}
\textbf{Input:} \textit{rawIUT}, \textit{PUT}, \textit{prioritizedMutant} \\
\textbf{Output:} \textit{FUT}, \textit{error} \\

\begin{algorithmic}[1]
    \STATE \textit{correctTest} $\leftarrow$ \textbf{false}
    \STATE \textit{FUT} $\leftarrow$ \textit{rawIUT}
    \STATE \textit{counter} $\leftarrow$ 0
    \WHILE{\textbf{not} \textit{correctTest} \textbf{and} \textit{counter} $<$ \textit{loopSize}}
            \STATE \textit{syntaxVerif}, \textit{errorSyntax} $\leftarrow$ \textbf{Compile}(\textit{FUT})
        \STATE \textit{behaviorVerif}, \textit{errorBehavior} $\leftarrow$ \textbf{RunTest}(\textit{FUT}, \textit{PUT})
        \IF{\textit{syntaxVerif} \textbf{and} \textit{behaviorVerif}}
            \STATE \textit{correctTest} $\leftarrow$ \textbf{true}
        \ELSE
            \STATE \textit{augmentedPrompt} $\leftarrow$ \textbf{CONCAT}(\textit{FUT}, \textit{PUT}, \textit{errorSyntax}, \textit{errorBehavior}, \textit{prioritizedMutant})
            \STATE \textit{FUT} $\leftarrow$ \textbf{LLM}(\textit{augmentedPrompt})
            \STATE \textit{counter} $\leftarrow$ \textit{counter} + 1
        \ENDIF
    \ENDWHILE
    \STATE \textbf{return} \textit{FUT}, \textit{correctTest}
\end{algorithmic}
\end{algorithm}

\subsection{Test Case Generation Module}
\textbf{Approach overview}: This section details the process of generating test cases as outlined in Algorithm \ref{alg:PRIMG}. Following an initial mutation testing phase, we identify and select the surviving mutants. The primary objective of the proposed approach is to develop unit tests capable of eliminating the maximum number of these surviving mutants. The process begins by selecting a specific mutant as the testing target, with the selection mechanism described in Section \ref{sec:Prioritization module}. The prioritized mutant is then used to generate an initial prompt for the large language model (LLM). Subsequently, the process enters a refinement phase, designed to ensure the correctness of the generated test cases, as elaborated in subsequent sections. If a test case is deemed correct, it is incorporated into the final test suite. Conversely, if the refinement process fails to produce a correct test case, the task is classified as problematic or one for which PRIMG is unable to generate an appropriate test case.
\subsubsection{Initial Prompt}
Large Language Models (LLMs) are adept at performing tasks they have been trained on. However, fine-tuning LLMs to adapt to a new task is computationally expensive. Additionally, certain LLMs, such as Codex\cite{chen2021evaluating}, excel in generating code but are closed-source, making fine-tuning for custom tasks infeasible.

Prompt-based learning \cite{liu2023pre, zhang2022repairing} offers an effective alternative for adapting LLMs to new tasks. A prompt is a combination of natural language and/or programming language context used as input to LLMs. Studies have demonstrated that incorporating natural language instructions as hints within prompts significantly enhances the ability of LLMs to perform novel tasks \cite{nashid2023retrieval} .

In our approach, the initial prompt consists of three key components:  
\begin{itemize}
    \item \textbf{The Program Under Test (PUT):} This represents the original program prior to mutation.  
    \item \textbf{The Mutant Code:} This introduces modifications to the original program, representing potential faults or variations.  
    \item \textbf{An Initial Test File:} This file is designed to validate the functionality and setup of the PUT within the testing framework. The test file is selected based on the following criteria:  
    \begin{enumerate}
        \item The test file should test functions defined in the PUT. This ensures the large language model (LLM) receives sufficient information about the testing environment and the imported libraries required to execute the tests.  
        \item The test file should contain diverse test cases. This enhances the LLM's understanding of the structure and variety of test cases, aiding in the generation of comprehensive and accurate tests.  
    \end{enumerate}
\end{itemize}
\subsubsection{Refining Process}
As illustrated in Algorithm \ref{alg:RefiningTest}, after generating the raw Initial Unit Test (IUT), we employ a refining process to enhance its quality and effectiveness. The refining process consists of two primary steps:

\paragraph{Syntax Verifier and Behavior Verifier}
In this step, we verify the correctness of the generated test case in two stages: 
\begin{enumerate}
    \item  \textbf{Syntax Verification}: The test case is compiled, and logs are reviewed for syntax errors.
    \item \textbf{Behavior Verification}: The test case is executed against the PUT. A valid test case must successfully pass when run on the PUT.
\end{enumerate}
\paragraph{Prompt Augmentation}
If the previous step fails (e.g., the test case fails to compile or the PUT fails when tested), we proceed to prompt augmentation. In this step:
\begin{itemize}
    \item A new prompt is constructed using the PUT, the mutant code, the FUT, and the error logs from the previous step.
    \item This enhanced prompt is used to generate a refined test case, referred to as the Final Unit Test (FUT), which resolves the issues encountered in the initial iteration.
\end{itemize}

By iteratively refining and augmenting the prompt, we ensure the generation of high-quality test cases capable of validating both the syntax and behavior of the program under test.

\subsection{Prioritization module }

\label{sec:Prioritization module}
\subsubsection{Dynamic Mutant Subsumption Graph}

In mutation testing, a test set \( T \) is considered \emph{mutation-adequate} with respect to a mutant set \( M \) if for every non-equivalent mutant \( m_i \in M \), there exists a test \( t \in T \) such that \( t \) detects \( m_i \)\cite{kurtz2014mutant}. However, mutation operators generate a significantly larger number of mutants than necessary, leading to redundancy. This redundancy is formally captured by the concept of \emph{minimal mutation}. A dominator set \( D \subseteq M \) is a minimal subset of \( M \) such that any test set adequate for \( D \) is also adequate for \( M \).
A key aspect of mutation testing is the \emph{subsumption relationship} between mutants. Given a finite set of mutants \( M \) and a finite test set \( T \), a mutant \( m_i \) is said to \emph{dynamically subsume} another mutant \( m_j \) if a test \( t \in T \) detects \( m_i \) and every test that detects \( m_i \) also detects \( m_j \). This relationship is \emph{strict} if \( m_i \) subsumes \( m_j \), but \( m_j \) does not subsume \( m_i \). If two mutants \( m_i \) and \( m_j \) are detected by exactly the same tests in \( T \), the subsumption is not strict \cite{kaufman2022prioritizing}.
The \emph{Dynamic Mutant Subsumption Graph (DMSG)} captures these subsumption relationships. In a DMSG, each node represents a maximal set of redundant mutants, and edges represent the subsumption relationships between sets. Specifically, if mutant \( m_i \) strictly subsumes mutant \( m_j \), there is a directed edge from the node containing \( m_i \) to the node containing \( m_j \). Any test that detects a mutant in the DMSG is guaranteed to detect all mutants subsumed by that mutant, including those in the same node or lower in the graph. This structure provides an efficient way to identify and eliminate redundant mutants, thereby improving the mutation testing process and reducing the computational burden on developers.
\subsubsection{Test Completeness Advancement Probability, TCAP} 
Test Completeness Advancement Probability (TCAP), as defined in \cite{kaufman2022prioritizing}, is employed to measure mutant usefulness in the context of test case generation. This measure quantifies the likelihood of a mutant, when presented as a test goal to a developer, eliciting a test case that enhances test completeness by killing the mutant and a maximum number of other mutants. This capture a key idea: a mutant is useful to be selected as the test goal, if when developing a test case that kills that mutant, other mutants are going to be killed by the same developed test case. This is due to the fact that the subsumed mutant, \( m_j \) is killed by the same tests that kills the dominant mutant \( m_i \). In a subsumption relationship, The test case that kills the dominant mutant are also killing the subsumed mutant.

As discussed in \cite{kaufman2022prioritizing}, we define the usefulness of a mutant in terms of its capacity to detect (or kill) other mutants when it is itself killed. 
However, for surviving mutants (mutants not killed by the initial test suite), it is not feasible to construct a Dominance Mutation Subsumption Graph (DMSG). Instead, TCAP exploits the structural information of the DMSGs of killed mutants. By analyzing these graphs, we can identify mutants that are likely to be useful as test goals. The intuition is that the test suite developed to kill these selected mutants will also effectively kill other mutants, thereby improving the overall efficiency of the test generation process and optimization the total number of needed tests. This approach builds on foundational ideas of mutation analysis and mutant subsumption as explored in prior research \cite{kurtz2014mutant, just2017inferring}. By leveraging the dominance relationships and subsumption hierarchies among mutants, we aim to streamline the test case generation process, improving both coverage and effectiveness throw the usage of a supervised machine learning model. 

The usefulness factor of a mutant is defined as  measurement of its ability to reveal additional mutants when used as a target for test case generation. The usefulness factor of a mutant has two properties: 
\begin{itemize}
    \item Dominator mutants, have usefulness factor equal to 1:
    By definition, each test that kills a dominator mutants kills every other subsumed mutant. 
    \item Subsumed mutants, have a usefulness factor less then 1 and greater then 0
\end{itemize}

\begin{figure*}[t!]
\centering
\includegraphics[width=.9\textwidth]{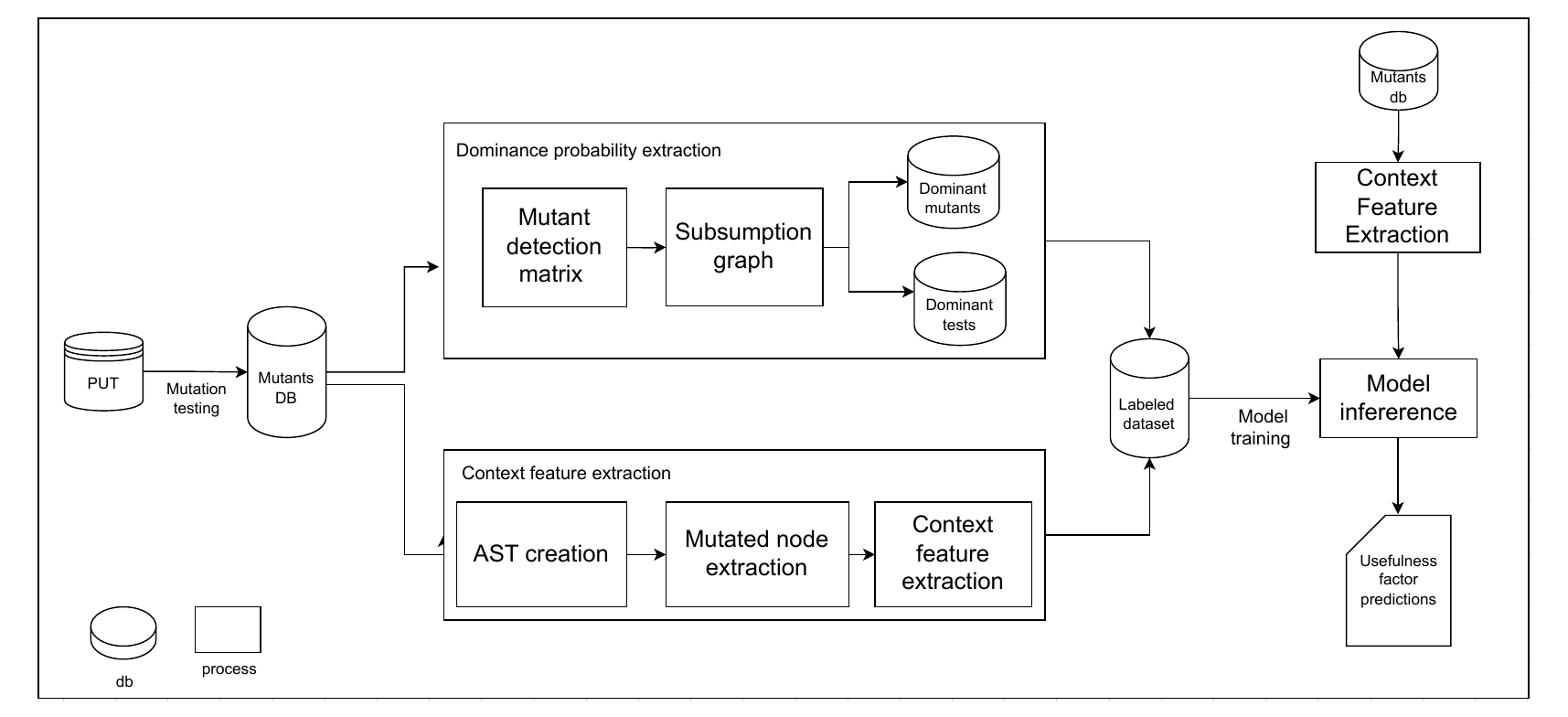}
\caption{Overview of the dataset labeling process}
\label{PM}
\Description{}
\end{figure*}
\subsubsection{Machine learning module}
\label{sec:MLM}
\begin{itemize}
\Description{}
\item \textbf{Context feature}: 
    
Building upon the modeling approach and the feature set proposed by Samuel J. Kaufman et al. \cite{kaufman2022prioritizing}, we incorporated program context features extracted from a program's Abstract Syntax Tree (AST). For a given mutated AST node, we examined its syntactic context, including parent and child nodes within the AST as well as nesting details, and its semantic context, such as the data types of expressions, operands, and method parameters.
It is important to highlight that previous studies\cite{zhang2016predictive} utilized information derived from test execution data, such as code coverage, to make predictions in the context of mutation analysis. As done in \cite{kaufman2022prioritizing}, our approach aims to identify mutants that are beneficial for generating test cases that have not yet been created, aligning with the goals of mutation testing.

\item \textbf{Feature extraction and dataset labeling}: Figure \ref{PM} shows an overview of our dataset labeling approach and model training. Since we can only construct the DMSG for the killed mutants, only the killed mutant will be used for training the model.

After an initial mutation testing campaign using the initial test suite of the PUT, we create our labeled dataset. The labeling of the dataset is composed of two modules: Dominance probability extraction and Context feature extraction.
The first component is the dominance probability extraction components which aims to label the killed mutants with an initial usefulness factor. Given the mutation testing results, we create the mutation matrix which maps test cases to mutants, indicating which test cases can detect (kill)  or not specific mutants. We use the mutation matrix to construct the DMSG of mutants relative to the initial test suite as defined in the previous section. We extract the parent node which contains the dominant mutants. Those identified dominant mutants are labeled as useful having the useful factor equal to 1. The other mutants are labeled as having a useful factor equal to zero.

The second component is context feature extraction module which aims to extract the context features from the Abstract Syntax Tree (AST) described in the previous section. First we start by comparing the AST of the mutant with the one of the PUT. We identify the mutated node. Using the mutated node we extract the remaining context features of the mutated node and its parent node context features.

using the usefulness factor and the context features, we create the labeled dataset that will be used for training and inference.
\item \textbf{Machine learning models}:
We evaluated a small number of model design choices and training settings against an intrinsic measure of model performance with the ultimate goal of choosing a single model for downstream evaluation. Each training setting was a kind of hold-one-out train-test split, using each held out project (or smart contract in a given project) at a time as the evaluation set. We trained all models using Scikit-Learn \cite{pedregosa2011scikit} and evaluated every combination of the following choices:

We have chosen these configuration: 
\begin{itemize}
        \item \textbf{Model choice}: We compared ridge regression to a random forest regression (max depth of 3 and 10 trees). \cite{kurtz2018improving, kaufman2022prioritizing} .
        \item \textbf{Features set}: We used a comprehensive set of features (All features, enumerated in \cite{just2017inferring}).
        \item \textbf{Training setting}: We evaluated the following train-test splits: 
        \begin{itemize}
            \item \textit{All Projects}: training on mutants from all projects, including those in non-held-out smart contracts in the same project, and testing on mutants in the held-out smart contracts;
            \item \textit{Project-Only}: training on mutants in all but one held-out smart contract in a single project and testing on mutants in the held-out smart contract.
        \end{itemize}
\end{itemize}

\end{itemize}
Building on findings in \cite{kaufman2022prioritizing}, we choose the following configuration for the experiments in the evaluation section:

\textit{Model=Ridge regression, Features=All, Training Setting=Project-Only}

\section{Evaluation}

In this section, we describe the evaluations we designed and conducted to investigate the following research questions.

\begin{enumerate}
\item \textbf{RQ1. How efficient is the pipeline for test case generation?}

\item \textbf{RQ2. How useful is the prioritization technique compared to random for efficient mutation testing?}
\end{enumerate}
\subsection{Experimental Setup}
In this section, we present our experiment setup. Specifically, we describe the mutant generation tool, clarify the LLM of PRIMG and its setup, and indicate the benchmark data sets used in our experiments.
We conducted the experiment on an EC2 instance of type g5.xlarge with 26 GiB of GPU memory.

\subsubsection{Experiment Parameters}

We call the refining process on the LLM up to 10 times and collect the output that meets two criteria: The generated test must compile correctly and then pass the PUT. If after 10 runs, the LLM is not able to generate an output that satisfies these two conditions, we consider the mutant as a problematic task or a task for which PRIMG is not able to generate a test case.

For the prioritization module, for each project we train a machine learning model to predict the survived mutant usefulness factor. As mentioned in section \ref{sec:MLM}, we use the configuration of: ridge regression model with all features. As we have already explained, the model is trained on mutants that have already been killed.

We use llama 3.1 with 8B parameter as the large language model for the generation of test case and refinement of test cases. Llama3.1 is the leading 

\subsubsection{Mutant Generator}

To apply mutation testing, we need to generate different mutant versions of a PUT by modifying different lines of code to introduce vulnerabilities. For this purpose
we use SuMo \cite{barboni2021sumo} as mutant generation tool. SuMo is a mutation testing tool adapted for smart contracts of Solidity. It uses different mutation operators to generate mutants including 25 Solidity-specific operators and 19 general-purpose mutation operators.

For testing, we used Ganache\footnote{Truffle Suite, \textit{Ganache - One Click Blockchain}, \url{https://trufflesuite.com/ganache/}, accessed on Nov 26, 2024.}
as a local blockchain network for testing and deploying smart contracts and executing transactions for mutants and PUT. We use hardhat as testing framework\footnote{Hardhat, \url{https://hardhat.org/}, accessed on Nov 26, 2024.}.
\subsubsection{Data sources}

\begin{table*}[ht]
\centering
\caption{Data sources}
\label{tab:DS}
\begin{tabular}{|c|c|c|c|}
\hline
 Project Id & \#  smart contracts & \# initial test files  & Initial mutation score \%    \\ \hline
 Allobase & 31 & 12 & 35.89  \\ \hline
 Particle & 19 & 7 & 46.86  \\ \hline
 Quadrata& 10 & 9 & 77  \\ \hline
\end{tabular}
\end{table*}

To conduct our experiments, we chose real world projects that we extracted from code4arena\footnote{\url{https://code4rena.com}}. Code4arena is a platform facilitating decentralized security audits of smart contracts through competitive audit contests conducted by a global community of security researchers. We selected three different projects. The details of the selected project is in the following Table \ref{tab:DS}.
To ensue rigor, we chose projects meeting these specific criteria:
\begin{itemize}
    \item Number of smart contracts: this criterion ensures the selection of complex projects by focusing on those with more than 10 smart contracts, avoiding overly straightforward examples.
    \item Initial test suite: in the selected project, the initial test suite must have at least one test case testing each smart contract.
    \item Mutation score: projects are chosen with an initial mutation score (MS) below 80\%, ensuring that the existing test suite is insufficient and requires additional test cases to improve coverage.
    
\end{itemize}

All scripts and dataset can be found here\footnote{\url{https://github.com/Salah-SH/llm-mutation-testing}}. 
\subsection{Empirical Study Results}

This section discusses our findings for each research question (RQ), focusing on the performance and efficiency of the proposed approaches.

\subsubsection{\textbf{Evaluation of the Performance of the Test Case Generation Module}}\hfill

In this subsection, we target to answer the first research question. We divide the RQ1 to two sub questions and answer each one of them presenting our findings.
\begin{tcolorbox}
\textbf{RQ1.} How efficient is the pipeline for test case generation?
\end{tcolorbox}

In this research question, we evaluate the performance of the test case generation module with respect to two key aspects: (1) the effectiveness of refining module in generating correct test cases and (2) the ability of these generated tests to kill new mutants that were not initially addressed by the original test suite. To ensure a comprehensive analysis, RQ1 is divided into the following subquestions:

\begin{itemize}
    \item \textbf{RQ1.1:} How efficient is the pipeline for generating syntactically and behaviorally correct tests?

    The objective here is to assess the accuracy of the test cases generated by the LLM. Correctness is evaluated on two dimensions: \textbf{syntactic correctness:} The test case compiles successfully, and \textbf{behavioral correctness:} The test case passes the original Program Under Test (PUT) without errors.

    \item \textbf{RQ1.2:} What is the impact of the size of the refining loop on the correctness of the generated tests?

    This sub-question investigates how the number of iterations in the refining loop impacts the quality of the generated test cases. Specifically, it examines whether iterative improvements significantly enhance the syntactic and behavioral correctness of the test cases.
    
    %
\end{itemize}

To answer RQ1.1, test cases were generated by prompting the LLM with and without the refining process using different loop sizes. A test case is considered correct if it successfully compiles and passes the original PUTs. The results were collected from three projects, as listed in Table \ref{tab:DS}. 

To answer RQ1.1 and statistically evaluate the impact of the refining process, we formulated the following hypotheses \cite{wohlin2012experimentation}:

\noindent\textbf{RQ1.1-H0:} \textit{The refining process does not significantly improve the number of correct tests compared to a single-shot prompt.}

\noindent\textbf{RQ1.1-H1:} \textit{The refining process significantly improves the number of correct tests compared to a single-shot prompt.}

\begin{figure}
\includegraphics[width=1\columnwidth]{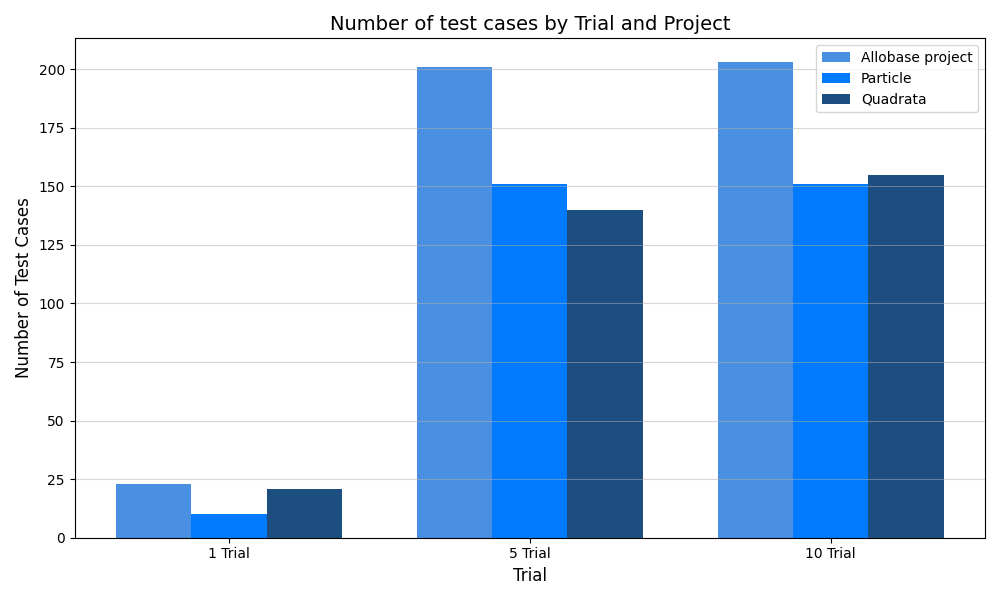}
\caption{Number of test cases by trial and projects }
\label{fig:refResults}
\Description{}
\end{figure}

To evaluate the efficiency of the test generation process, we analyzed the performance of test case generation with respect to the number of correctly generated test cases across varying trial counts. Table \ref{fig:refResults} summarizes the results obtained when running the process with 1 trial, 5 trials, and 10 trials. To assess the statistical significance of the differences between these configurations, we performed pairwise Z-tests for proportions, a statistical method suitable for comparing success rates across independent trials. 

The comparisons were conducted across three projects: Allobase, Particle, and Quadrata. The results of the Z-tests for each project are as follows: 
\begin{itemize} 
    \item \textbf{Allobase:} Significant differences were observed between 1 trial and 5 trials ($Z = -19.08, p < 0.05$) as well as 1 trial and 10 trials ($Z = -19.20, p < 0.05$). No significant difference was found between 5 trials and 10 trials ($Z = -0.22, p > 0.05$). 
    \item \textbf{Particle:} Significant differences were observed between 1 trial and 5 trials ($Z = -12.13, p < 0.05$) and 1 trial and 10 trials ($Z = -12.13, p < 0.05$). No significant difference was found between 5 trials and 10 trials ($Z = 0.0, p = 1.0$). 
    \item \textbf{Quadrata:} Significant differences were observed between 1 trial and 5 trials ($Z = -10.24, p < 0.05$) and 1 trial and 10 trials ($Z = -10.54, p < 0.05$). No significant difference was found between 5 trials and 10 trials ($Z = -0.35, p = 0.73$). 
\end{itemize}

\begin{table}[h!]
\centering
\resizebox{\columnwidth}{!}{
\begin{tabular}{|c|c|c|c|}
\hline
\textbf{Project} & \parbox{2cm}{\centering \textbf{Single Shot} } & \parbox{2cm}{\centering \textbf{Refining process} \\ \textbf{loop of 5}} & \parbox{2cm}{\centering \textbf{Refining process} \\ \textbf{loop of 10}} \\ \hline
\textbf{Allobase} &23/500 & 201/500 & 203/500 \\ \hline
\textbf{Particle } & 10/500 & 151/500 & 151/500 \\ \hline
\textbf{Quadrata} & 21/500 & 140/500 & 145/500 \\ \hline
\end{tabular}
}
\caption{Number of correct test case generated in different Projects using different loop size}
\label{tab:project_results}
\end{table}

Specifically, for the projects Allobase, Particle, and Quadrata, the Z-test results provide strong evidence that increasing the number of trials significantly enhances the success rates of generating correct tests. These findings suggest that using more trials contributes to better test generation performance, as evidenced by the statistically significant improvements observed when increasing from 1 trial to 5 trials and from 1 trial to 10 trials. However, no significant differences were found between 5 and 10 trials across all projects, indicating diminishing returns beyond 5 trials. The observed differences in success rates are statistically significant under the experimental conditions and datasets analyzed. Thus, increasing the number of trials, particularly from 1 to 5, outperforms single-shot attempts in generating correct tests, highlighting the value of incorporating multiple trials in improving LLM-generated test cases.

Additionally, as shown in Table \ref{tab:project_results}, the impact of the refining process on the performance of test case generation is evident. In the \textit{Allobase} project, the single-shot approach generates only 4.6\% of correct tests, whereas using the refining process with a loop size of 5 iterations results in a substantial improvement, with 40.2\% of tests being correct. Similarly, for the \textit{Particle} project, the single-shot approach yields only 2\% correct tests, while applying the refining process with 5 iterations increases the correct tests to 30.2\%. In the case of \textit{Quadrata}, the single-shot approach results in 4.2\% correct tests, and refining with 5 iterations improves this to 28\%. These results indicate a significant performance boost when employing the refining module, demonstrating that iterative refinement is a crucial factor in improving the accuracy and effectiveness of the test case generation module. Therefore, we conclude that utilizing the refining module greatly enhances the performance of the test case generation process, making it a valuable strategy for generating higher-quality test cases in all evaluated projects.

\begin{tcolorbox}
\textbf{Finding 1:} The refining module can significantly improve the correctness of the generated tests.
\end{tcolorbox}

To answer \textbf{RQ1.2}, we conducted an additional analysis focusing on the effect of loop size on the correctness of the generated tests. In this research question, our primary objective is to evaluate the role of the number of iterations (loop size) within the refining module on the overall performance of the test case generation process. The hypothesis guiding this investigation was based on the premise that increasing the number of loops could lead to more refined and accurate test cases. The results of this analysis, including detailed performance comparisons for each project, are presented in Table \ref{tab:project_results}.

We formulated the null hypothesis (H0) and alternative hypothesis (H1) to test the impact of the number of trials as follows:

\noindent\textbf{RQ1.2-H0:} \textit{Both 5 and 10 loops significantly improve the number of correct test cases compared to single-shot for all projects.}

\noindent\textbf{RQ1.2-H1:} \textit{Increasing the number of trials in the test generation process positively impacts the success rate of generated test cases.}

To analyze the performance of the test generation module across different trial configurations (1 trial, 5 trials, and 10 trials), we employed pairwise Z-tests for proportions. These tests compare the success rates of correctly generated tests for different numbers of trials. The Z-test was applied to all three projects, Allobase, Particle, and Quadrata, and the results are summarized below:

\begin{itemize}
    \item \textbf{Allobase:} No significant difference was found between 5 and 10 trials ($Z = -0.22, p > 0.05$).
    \item \textbf{Particle:} No significant difference was found between 5 and 10 trials ($Z = 0.00, p > 0.05$).
    \item \textbf{Quadrata:} No significant difference was found between 5 and 10 trials ($Z = -0.35, p > 0.05$).
\end{itemize}

These findings suggest that increasing the number of trials from 1 to 5 significantly improves the success rates of test case generation across all projects. However, further increases beyond 5 trials do not result in statistically significant improvements. The Z-test results emphasize the value of employing multiple trials, particularly up to 5, to enhance the accuracy and reliability of LLM-generated test cases.

\begin{tcolorbox}
\textbf{Finding 2:} No significant improvement is observed when increasing from 5 to 10 loops, suggesting diminishing returns beyond 5 iterations.
\end{tcolorbox}

Overall, the results suggest that the the refining loop technique using a loop size of 5 is the best configuration for optimal test case generation performance.

\subsubsection{\textbf{Evaluation of the performance of prioritization module}}

In this subsection, we target the second research question investigating the impact and the effectiveness of the prioritization module for improving the efficiency of the generated tests.
\begin{tcolorbox}
\textbf{RQ2.} How useful is the prioritization technique compared to random for efficient mutation testing?
\end{tcolorbox}

To evaluate the prioritization module, we perform analysis on the tests generated for the projects as mentioned in Section \ref{table:overlaid}. For each project among the three projects, we generated 500 tests using the test case generation module.
We train a ridge regression model for each project as described in Section \ref{sec:Prioritization module}. We use the configuration \textit{Model=ridge regression, Features=all, Training Setting=Project-Only}.

We distinguish between 2 types of tests: (1) Prioritized test is a test case that have been generated targeting a prioritized mutant and (2) random test is a test that has been generated targeting a random mutant.

In Table ~\ref{table:overlaid} we illustrate the results of the mutation testing we executed using three random set of tests composed of 50 tests and one prioritized set of tests.
To answer RQ2, we conducted a hypothesis test on the results project Allobase, Particle and Quadrata as presented in Table ~\ref{table:overlaid}. We formulated the null hypothesis (H0) and the alternative hypothesis (H1) as follows:

\noindent\textbf{RQ2-H0:} \textit{There is no statistically significant difference in the performance of prioritized generated tests compared to randomly generated tests}

\noindent\textbf{RQ2-H1:} \textit{There is a statistically significant difference in the performance of prioritized generated tests compared to randomly generated tests}

Similar to RQ1.1 and RQ1.2, we compare the efficacy of these strategies (i.e., prioritized tests vs. randomly selected tests) by using Z-test.

\begin{itemize}
    \item \textbf{Allobase:} The Z-test for this comparison yielded $Z = -15.98, p < 0.05$, indicating a statistically significant difference between the two methods, with Method 1 outperforming Method 2.
    \item \textbf{Particle:} The Z-test for this comparison yielded $Z = -12.94$, $p < 0.05$, also showing a statistically significant difference between the two methods, with Method 1 performing better than Method 2.
    \item \textbf{Quadrata:} The Z-test for this comparison yielded $Z = -14.12, p < 0.05$, demonstrating a statistically significant difference between the two methods, with Method 1 outperforming Method 2.
\end{itemize}

These results suggest that, across all three projects, Allobase, Particle, and Quadrata, the test suite composed of prioritized tests significantly outperforms the test suite composed of random tests set generated using random selection of mutants in killing mutants. The observed differences in success rates are statistically significant under the experimental conditions analyzed, highlighting the effectiveness of the prioritization technique in killing mutants.
Thus, the prioritization module consistently enhances test performance, demonstrating its superiority over random selection in all the evaluated projects. These results highlight the potential of the prioritization strategy as a valuable tool in mutation testing, contributing to more efficient and targeted test generation in software testing practices.

\begin{tcolorbox}
\textbf{Finding 3:} There is a statistically significant difference in the performance of prioritized generated tests compared to randomly generated tests in killing mutants.
\end{tcolorbox}

\begin{table*}[h!]
\centering
\begin{tabularx}{\textwidth}{|X|X|X|X|X|X|}
\hline
\multicolumn{1}{|c|}{\textbf{Project}} & 
\multicolumn{3}{c|}{\textbf{Randomly Selected Tests}} & 
\multicolumn{1}{c|}{\textbf{Prioritized Tests}} & 
\multicolumn{1}{c|}{\textbf{Ground Truth}} \\ 
\cline{2-4}
 & \textbf{1} & \textbf{2} & \textbf{3} &  &  \\ 
\hline
Allobase  & 200 & 158 & 8 & 300 & 465 \\ 
\hline
Particle  & 196 & 54 & 180 & 250 & 350 \\ 
\hline
Quadrata & 55 & 205 & 200 & 256 & 403 \\ 
\hline
\end{tabularx}
\caption{Overlaid Table representing the number of killed mutants using Randomly Selected Tests, Prioritized Test, and Ground Truth}
\label{table:overlaid}
\end{table*}

\section{Discussions}

\subsection{Interpretation of the importance of the loop size on the effectiveness of test case generation}
Although the refining module have shown impressive results in generating and refining generated tests, experiments in the evaluation section shows that increasing the size of the loop does not considerably improve the number of correctly generated tests since even having a larger trial number, the LLM still fails to generate correct tests no matter the number of repetitions. In fact,  LLMs operate based on statistical patterns in their training data rather than true understanding of programming concepts or software semantics\cite{dou2024s}. This can lead to what researchers call "hallucination" - generating plausible-looking but incorrect code that doesn't actually fix the underlying bug\cite{hossain2024deep}. As Dou et al. observed in their extensive study, functional bugs constitute the highest proportion of errors in LLM-generated code fixes, with misunderstanding and logic errors being particularly prevalent\cite{dou2024s}. Even after multiple attempts, an LLM may continue to make similar mistakes if it fundamentally misunderstands the problem or lacks the specific knowledge needed to generate a correct fix.
Besides, the token-by-token generation process of LLMs can lead to inconsistencies in longer code sequences, even if individual parts seem correct. This is particularly problematic for bug fixes that require changes across multiple lines or functions. The authors in \cite{dou2024s} observed that LLMs tend to perform better on localized, single-line fixes but struggle with bugs that require broader context or multiple coordinated changes. Furthermore, the limited context window of most LLMs can prevent them from fully considering all relevant parts of a large codebase, leading to fixes that may solve one issue but introduce new bugs elsewhere. These structural limitations of LLMs mean that even with repeated attempts, they may struggle to generate consistently correct and comprehensive bug fixes for more complex software issues.

\subsection{Adequacy of our approach for practical applications}

The PRIMG framework presents a practical and efficient solution for test suite generation by integrating automated test case generation with an advanced prioritization module. Leveraging Large Language Models (LLMs), the framework automates the creation of diverse, high-quality test cases, minimizing the manual effort required from developers while ensuring robust coverage of edge cases and complex scenarios. Complementing this, the prioritization module employs machine learning to rank and focus on the most impactful mutants, optimizing the testing process by reducing redundant efforts and maintaining high mutation coverage with a smaller, more efficient test suite. This dual approach significantly enhances the efficiency of mutation testing, streamlining the traditionally resource-intensive process and enabling developers to concentrate on critical tests. Particularly in domains like smart contract development, where rapid and reliable testing is crucial, PRIMG demonstrates its value by reducing computational overhead and improving the precision and practicality of the testing workflow. Its combination of automation, prioritization, and scalability establishes PRIMG as a powerful tool for advancing software quality in both research and industrial contexts.

\section{Data Availability}
All the scripts and results that have been used during our experiments can be found in this github repository \url{https://github.com/Salah-SH/llm-mutation-testing}.

\section{Threat to Validity}

In this study, we employed two different prompt-based learning techniques: one-shot and prompting augmenting technique. However, we did not explore the potential impact of altering the natural language instructions or demonstrative examples (for the prompt augmenting technique) within our prompts. Modifying these instructions or utilizing different demonstrative examples more closely aligned with the PUT’s functionality could potentially enhance the results. 

A potential threat to the validity of our study arises from the choice of mutation operators defined within the SuMo framework\cite{BARBONI2022111445}. Mutation operators play a critical role in generating test cases, and the limited set of operators available in SuMo may not adequately represent all possible vulnerabilities in smart contracts. This can lead to a biased set of mutants, limiting the effectiveness of the generated tests in detecting a broad range of issues. If the operators do not cover diverse attack vectors or edge cases, the results of the mutation testing may not fully reflect the real-world security risks present in smart contracts, potentially compromising the generalize-ability of our findings.

Another validity threat arises from the exclusive use of Hardhat as the testing framework. While Hardhat is a widely adopted and robust tool for smart contract development and testing, its exclusive use may restrict the applicability of our results to other testing environments. Variations in how different frameworks execute and interact with smart contracts could lead to discrepancies in test outcomes. As a result, the conclusions drawn from Hardhat-based testing might not extend to projects using alternative frameworks, thus reducing the external validity of our study.

Additionally, this study did not evaluate the number of newly killed mutants or the impact of the new tests on the overall mutation score. Although this metric is important, our primary focus was on generating correct and valid test cases.

A further threat to validity arises from the practical challenges and resource limitations encountered during testing. While there is no inherent limitation in the availability of projects, the complexities involved in setting up the testing environment and the constrained resources for conducting extensive experiments restricted the scope of our analysis. Overcoming these challenges in future work could enable a broader evaluation across more diverse projects and environments.

\section{Conclusion}
This paper presented PRIMG (Prioritization and Refinement Integrated Mutation-driven Generation), a novel framework for incremental and adaptive test case generation tailored for Solidity smart contracts. By integrating a mutation prioritization module with a test case generation module leveraging Large Language Models (LLMs), PRIMG addresses key challenges in mutation testing, such as excessive test suite sizes and inefficiencies in test generation. The mutation prioritization module employs machine learning models trained on mutant subsumption graphs to predict the usefulness of mutants, enabling the selection of high-impact test goals. Simultaneously, the test case generation module iteratively refines LLM-generated tests to ensure syntactic and behavioral correctness.

Our evaluation on real-world Solidity projects demonstrated PRIMG’s effectiveness in improving mutation scores while reducing test suite size and computational overhead. The prioritization module consistently outperformed random mutant selection by enabling more efficient and targeted test case generation. Moreover, the refining process successfully mitigated common issues associated with LLM-generated tests, such as syntactic errors and limited bug-detection capabilities.

Although the current version of PRIMG employs mutant prioritization and LLMs to generate test cases for Solidity contracts programs, its design and evaluation methodology are fundamentally adaptable to various programming languages and models.
Therefore, as future work, it can be easily expanded to encompass other programming languages or incorporate new LLMs.

\bibliographystyle{ACM-Reference-Format}
\bibliography{main}


\end{document}